\documentclass[12pt]{article}
\usepackage{graphicx}
\usepackage{amsmath}
\usepackage{longtable}

\begin{document}

\makeatletter
\renewcommand*{\@cite}[2]{{#2}}
\renewcommand*{\@biblabel}[1]{#1.\hfill}
\makeatother

\title{Spatial Distribution and Kinematics of OB Stars}
\author{G.~A.~Gontcharov\thanks{E-mail: georgegontcharov@yahoo.com}}

\maketitle

Pulkovo Astronomical Observatory, Russian Academy of Sciences, Pul\-kov\-skoe sh. 65, St. Petersburg, 196140 Russia


The sample of 37 485 suspected OB stars selected by Gontcharov (2008) from the Tycho-2 catalogue has
been cleaned of the stars that are not of spectral types OV--A0V. For this purpose,
the apparent magnitude $V_T$ from Tycho-2, the absolute magnitude $M_{V_T}$ calibrated as a function of the
dereddened color index $(B_T-V_T)_0$, the interstellar extinction $A_{V_T}$ calculated from the 3D analytical
model by Gontcharov (2009) as a function of the Galactic coordinates, and the photometric distance
$r_{ph}$ calculated as a function of $V_T$, $M_{V_T}$, and $A_{V_T}$ have been reconciled in an iterative process. The
20 514 stars that passed the iterations have $(B_T-V_T)_0<0$ and $M_{V_T}>-5$ and are considered as a
sample of OV--A0V stars complete within 350 pc of the Sun. Based on the theoretical relation between
the dereddened color and age of the stars, the derived sample has been divided into three subsamples:
$(B_T-V_T)_0<-0.2^m$, $-0.2^m<(B_T-V_T)_0<-0.1^m$, and $-0.1^m<(B_T-V_T)_0<0^m$, younger than 100, $100-200$,
and $200-400$ Myr, respectively. The spatial distribution of all 20 514 stars and the kinematics analyzed
for more than 1500 stars with radial velocities from the PCRV and RAVE catalogues are different for
the subsamples, showing smooth rotations, shears, and deformations of the layer of gas producing stars
with the formation of the Gould Belt, the Great Tunnel, the Local Bubble, and other structures within
the last 200 Myr. The detected temporal variations of the velocity dispersions, solar motion components,
Ogorodnikov--Milne model parameters, and Oort constants are significant, agree with the results of other
authors, and show that it is meaningless to calculate the kinematic parameters for samples of stars with
uncertain ages or with a wide range of ages.

\newpage
\section*{INTRODUCTION}

Stars of spectral types O and B are interesting
primarily as young high-luminosity stars. The youth,
i.e., the short lifetime near the main sequence (MS),
is explained by the large mass of OB stars. They exist
only in regions of current or recent star formation, for
example, in Galactic spiral arms, and, therefore, are
good tracers of young and short-lived structures as
well as the spatial distribution of the gas from which
they have recently been formed. The kinematics
retained during their short lifetime is important for
understanding the evolution of the Galaxy.

OB stars are convenient for selection, because
they have approximately the same metallicity and,
consequently, a unique theoretical relation between
their mass, dereddened color, absolute magnitude,
and age. To illustrate this relation, Fig. 1 shows some
evolutionary tracks and isochrones on a Hertzsprung--Russell
(H--R) diagram of the form ``color $(B_{T}-V_{T})$ -- absolute magnitude $M_{V_T}$'' (here and below, all
tracks and isochrones were taken from the database
in Padova (http://stev.oapd.inaf.it/cmd; Bertelli et al.
2008; Marigo et al. 2008); $B_T$ and $V_T$ are the magnitudes
from the Tycho-2 catalogue (H\o g et al. 2000).

In the figure, the theoretical isochrones for solar metallicity
stars with ages of 1, 100, 200, 300, and
400 Myr are indicated by five solid black lines from
left to right, respectively. The diamonds with the
gray curve passing through them and the squares
indicate the evolutionary tracks for a solar-metallicity
star with a mass of 2.9 $M_{\odot}$ (on the zero-age main
sequence (ZAMS), this is a B7V star) and a star with
a mass of 2 $M_{\odot}$ (A0V on the ZAMS), respectively.
We see that in this part of the diagram the tracks and
isochrones are far from coincidence and the selection
of stars larger than some mass is not the same as the
selection of stars younger than some age. On the
other hand, we see that the B and A stars separate
rather clearly near $(B_{T}-V_{T})_{0}=0$. This means that,
without oversimplifying, stars with a mass on the MS
greater than 2.5 $M_{\odot}$ can be said to be high-luminosity
stars younger than 400 Myr; they are also OB stars
(including a small fraction of subgiants, giants, and
supergiants) and, in the absence of extinction, have
$(B_{T}-V_{T})_{0}<-0.05$ near the MS. Given the accuracy
of the photometry, $\sigma((B_{T}-V_{T})_{0})\approx0.05$, below
we select stars with $(B_{T}-V_{T})_{0}<0$ for the sample to
be complete.

It can also be seen from the figure that, in contrast
to the A stars, the OB stars rapidly change their
color to red after core hydrogen exhaustion, i.e., at
the subgiant stage. Therefore, theoretically, the upper
right quarter of the shown diagram must be devoid of
stars.

Thus, the task is seemingly simple -- the selection
of OB stars as the bluest stars. However, some of the
low-mass stars at the core helium burning stage (hot
subdwarfs, blue horizontal-branch halo and thickdisk
stars) (Gontcharov et al. 2011) and extremely
fast OB stars (Tetzlaff et al. 2011) have the same color
and absolute magnitude. They can be revealed only
by their spectroscopy and kinematics. In addition,
even a slight extinction leads to such a reddening of
an OB star that it has color indices approximately as
those for stars of type A or even later.

Gontcharov (2008a) showed that the set of five
photometric bands, $B_T$ and $V_T$ from Tycho-2 and $J$,
$H$, and $Ks$ from the 2MASS catalogue (Skrutskie
et al. 2006), is sufficient for the separation of OB and
later-type stars on color–color $(V_{T}-H)$ - $(J-Ks)$
diagrams, although some admixture of extraneous
stars remains. In this case, the spectral classification
of stars is not used at all. All stars in the region
of the color--color diagram determined relative to the
ZAMS and the theoretical reddening line for B5 stars
are deemed OB stars. As a result, a sample of 37 485
suspected OB stars from the Tycho-2 catalogue was
produced. Comparison with the well-known spectral
classification from the Hipparcos Input Catalogue
(HIC, Turon et al. 1993) and Tycho-2 Spectral Types
(TST) catalog (Wright et al. 2003) confirms that the
selected stars are of types O–B with an admixture
of later types. This admixture forces the sample by
Gontcharov (2008a) to be considered as a preliminary
one.

Smaller but cleaner samples of OB stars were
produced on its basis \emph{by invoking the well-known
spectral classification}: a sample of 15 670 stars
to construct the 3D analytical extinction model
(Gontcharov 2009) and a sample of 11 990 stars to
investigate the spatial variations of the extinction
coefficient $R_V$ (Gontcharov 2012a). However, as was
shown by Gontcharov (2008a), using the well-known
spectral classification causes strong observational
selection and does not allow the spatial distribution
of OB stars to be investigated. In addition, it can be
seen from our comparison of the sample sizes (37 485
stars against 15 670 and 11 990) that doubts about
whether the stars belong to OB ones arise for most
stars of the preliminary sample.

For the stars of the preliminary sample with accurate
parallaxes $\pi$ and proper motions $\mu$ from the Hipparcos
catalogue (ESA 1997), Gontcharov (2008a)
found the correlation $M_{V_{T}}=0.45M'_{V_{T}}-1$ between
the absolute magnitude $M_{V{T}}$ and the reduced proper
motion $M'_{V_{T}}=V_{T}-A_{V_{T}}+5+5\cdot~\lg(\mu)$, where $A_{V_{T}}$
is the extinction in the $V_T$ band estimated by Gontcharov
(2008a) for each star from its color index, $\mu=(\mu_{\alpha}\cos\delta^2+\mu_{\delta}^2)^{1/2}$
is the total proper motion in arcseconds.
This correlation is explained by the correlation
between $\pi$ and $\mu$ in the absence or consideration
of systematic motions of the stars relative to the Sun
and in the absence of observational selection. For
example, Gontcharov (2008a) took into account the
Galactic rotation and the solar motion to the apex. As
a result, for each star of the sample we can calculate
the distance
\begin{equation}
\label{rrpm}
r_{rpm}=10^{(V_{T}-A_{V_{T}}-(0.45M'_{V_{T}}-1)+5)/5},
\end{equation}
Gontcharov (2008a) called it the photometric one,
although here and below it would be more correct to
call it the photoastrometric one, following the authors
of the Besancon model of the Galaxy (BMG, Robin
et al. 2003), as distinct from the classical definition of
the photometric distance via the calibration of $M_{V_{T}}$
as a function of $((B_{T}-V_{T})_{0})$. Such a calibration
for the stars of the preliminary sample is performed
below. This allowed us to calculate the photometric
distances
\begin{equation}
\label{rph}
r_{ph}=10^{(V_{T}-A_{V_{T}}-M_{V_{T}}+5)/5}, M_{V_{T}}=f((B_{T}-V_{T})_{0})
\end{equation}
where $f$ is the function found below.

Using $r_{rpm}$ for OB stars is restricted by the fact
that because of their youth, the assumption about the
absence of systematic motions is invalid and no accurate
allowance for these motions is possible. Therefore,
Gontcharov (2008a) reached only general conclusions
about the spatial distribution and kinematics
of OB stars. This paper is devoted to their detailed
analysis based on $r_{ph}$ and not on $r_{rpm}$. We calculate
$r_{ph}$, eliminate the admixtures of extraneous stars,
divide the sample into subsamples of stars with different
ages, and study in detail their spatial distribution
and kinematics.

\section*{PHOTOMETRIC DISTANCES}

Let us estimate the accuracies of the quantities
in Eq. (2). $V_T$ is known from observations with an
accuracy that, as a rule, is better than $0.1^{m}$. $M_{V_{T}}$ depends
on $(B_{T}-V_{T})_{0}$. As Perryman (2009) showed,
the scatter of individual values of $M_{V_{T}}$ around this
dependence for stars of the same spectral subtype
and luminosity class is usually larger than $0.5^{m}$. In
turn, $(B_{T}-V_{T})_{0}$ depends on $B_T$ , $V_T$, and $A_{V_{T}}$. The
extinction $A_{V_{T}}$ can be determined by various methods,
but we imply either a certain spectral energy
distribution for a given star in the absence of extinction
or a certain mean extinction in a given region
of space. The accuracy of calculating the extinction
$\sigma(A_{V_{T}})$ for a specific star is determined by the scatter
of the spectral energy distributions for stars of the
same type in the former case and by the scatter of
the individual extinctions for neighboring stars in the
latter case. As was shown by Gontcharov (2011),
$\sigma(A_{V_{T}})\approx0.3^{m}$ within the nearest kiloparsec in both
cases. Consequently, the accuracy of determining $r_{ph}$
from Eq. (2) depends primarily on the accuracy of determining $M_{V_{T}}$. Therefore, the method of determining
$A_{V_{T}}$ is not so important and we can relate all quantities
in Eq. (2) cyclically by admitting the dependence
of $A_{V_{T}}$ on $r_{ph}$ (as well as on $l$ and $b$) according to the
3D analytical extinction model (Gontcharov 2009)
by taking into account the relation $A_{V_{T}}=1.1A_{V}$ in
accordance with the most commonly used extinction
laws, for example, from Rieke and Lebofsky (1985).
When calculating the parameters of the extinction
model, Gontcharov (2009) used both OB stars and
stars of other types (Gontcharov 2012b). Therefore,
the model is valid for all stars.

Thus, using the 3D model to estimate $A_{V_{T}}$ allows
$M_{V_{T}}$, $r_{ph}$, and $A_{V_{T}}$ for most stars of the preliminary
sample to be reconciled in an iterative process, with
the minority of extraneous stars having been thrown
away.

\subsection*{$M_{V_{T}}=f((B_{T}-V_{T})_{0})$ Calibration}

When calibrating $M_{V_{T}}$ as a function of $(B_{T}-V_{T})_{0}$
\begin{equation}
\label{mvt}
M_{V_{T}}=V_{T}+5-5\lg(r_{HIP})-A_{V_{T}},
\end{equation}
where $A_{V_{T}}=1.1A_{V}$, and $A_{V}$ was calculated from the
mentioned 3D extinction model
\begin{equation}
\label{btvt0}
(B_{T}-V_{T})_{0}=(B_{T}-V_{T})-E_{(B_{T}-V_{T})},
\end{equation}
where the reddening $E_{(B_{T}-V_{T})}$ was also calculated
from the 3D extinction model and the 3D map of
$R_V$ variations (Gontcharov 2012a): $E_{(B_{T}-V_{T})}=1.1A_{V}/R_{V}$.
For the calibration, we used all 3237
stars of the preliminary sample with parallaxes $\pi>5$ mas from the new version of Hipparcos (van
Leeuwen 2007), provided that the relative accuracy of
the parallax $\sigma(\pi)/\pi<0.2$. The limitations stem from
the fact that the Lutz--Kelker and Malmquist biases
(Lutz and Kelker 1973; Perryman 2009, pp. 209--211) did not manifest themselves only for the stars
nearer than 200 pc.

Figure 1a shows the positions of 3237 calibration
stars on the $(B_{T}-V_{T})$ -- $M_{V_{T}}$ diagram, where
the color was not dereddened. Figure 1b shows the
positions of the same stars on the $(B_{T}-V_{T})_{0}$ -- $M_{V_{T}}$
diagram, i.e., after dereddening. We see that the overwhelming
majority of stars are placed by the applied
extinction model in the expected region between the
1-Myr isochrone and the ``loops'' on the isochrones
corresponding to the onset of nuclear reactions in the
shells of the stars when they become subgiants and
swiftly redden. We see a considerable number of stars
rightward of the 400-Myr isochrone, i.e., A stars.

The dashed line in Fig. 1b indicates the adopted
calibration
\begin{equation}
\label{calib}
\begin{split}
M_{V_{T}}=&-20.749(B_{T}-V_{T})_{0}^{4}-{}+22.376(B_{T}-V_{T})_{0}^{3}-\\
&-7.2134(B_{T}-V_{T})_{0}^{2}+6.3192(B_{T}-V_{T})_{0}+1.32.
\end{split}
\end{equation}

It is semiempirical, reflecting a different fraction of
stars deviated from the ZAMS (the 1-Myr isochrone)
for different $(B_{T}-V_{T})_{0}$. Only in the central part of
this figure is this fraction large and the calibration deviates
noticeably from the ZAMS. The scatter of the
3237 stars under consideration relative to this curve
is $0.8^{m}$. This gives $r_{ph}$ with a relative accuracy of
40\%. As we see from Fig. 1, this accuracy cannot be
higher, because it is determined by the $M_{V_{T}}$ difference
between the 1-Myr isochrone (ZAMS) and the loops
of the isochrones (the making of a subgiant).

The distances $r_{ph}$ are important in our study,
because the preliminary sample and the sample
produced below contain only 9926 (26\%) and 5605
(27\%) Hipparcos stars, respectively, with the accuracy
of their $r_{HIP}$ being higher than 40\%. Thus,
for the overwhelming majority of the stars under
consideration, $r_{HIP}$ are either lacking or less accurate
than $r_{ph}$.

\subsection*{Reconciling $M_{V_{T}}$, $r_{ph}$ and $A_{V_{T}}$}

Below, we perform the main procedure of our
study -- a mutually reconciled refinement of $M_{V_{T}}$, $r_{ph}$ and $A_{V_{T}}$
by iterations, given the mentioned dependences
of these quantities on one another. Since
the variations and errors in $A_{V_{T}}$ are small compared
to those for $M_{V_{T}}$ and $r_{ph}$, the calculations converge
with confidence and reach a relative computational
accuracy of 0.001 no later than the twentieth iteration
for more than 90\% of the stars.

The iterations did not converge for giants, supergiants,
shell stars, peculiar and late-type stars,
because the calibration (5) is invalid for them. One
of the advantages of this approach is that no stars are
excluded from the sample in advance.

Out of the stars that passed the iterations, 20 514
have $(B_{T}-V_{T})_{0}<0$ and $M_{V_{T}}>-5$ (the second
condition rejects the OB giants and supergiants that
passed the iterations, but the calibration (5) gives
unrealistic values for them). They are considered
below in this study as the final sample of OB stars
near the MS (presumably with an age of less than
400 Myr). Out of these stars, 7526 (37\%) are from
Hipparcos and 15 734 (77\%) have the spectral classification
from TST or HIC. Among the classified stars,
the distribution in spectral types is the following: O --
73 (0.5\%), B -- 8494 (54\%), A0 -- 6165 (39\%) A1-A9 -- 944 (6\%), F -- 32 (0.2\%), G -- 9, K -- 17.
Thus,
14 732 stars (94\% of the classified stars) are of types
O-A0, which allows the selection of OB stars to be
recognized as successful (some fraction of A stars is
present, but it will be useful below for comparison
when dividing the sample into age subsamples).

The distances $r_{HIP}$, $r_{rpm}$ and $r_{ph}$ agree within
350 pc of the Sun; further out, systematic effects have
an impact. At great distances, $r_{ph}$ seem to be most
accurate.

\section*{SPATIAL DISTRIBUTION}

Figure 2 shows the distribution of the sample stars
in projection onto the $XY$, $XZ$, and $YZ$ planes (the
$X$ axis is directed toward the Galactic center, the $Y$
axis is in the direction of rotation, the $Z$ axis is directed
toward the North Pole, the Sun is at the center,
the distances are in kpc): (a) with $r_{HIP}$ for 7288 Hipparcos
stars with $\pi>0$ mas, (b) with $r_{rpm}$ for all
20 514 stars, (c) with $r_{ph}$ for the same stars. We see
that the sample from Tycho-2 encloses an appreciably
larger space than does the sample from Hipparcos.
The Great Tunnel passing from the second octant to
the sixth one (Welsh 1991) is clearly seen on the $XY$
plots as a region of low star density near the Sun
bounded by concentrations of stars; the Gould Belt
(Perryman 2009, pp. 324--328; Gontcharov 2009) is
seen on the $XZ$ plots as a region of enhanced star
density inclined to the equator approximately by 17$^{\circ}$;
the radial (relative to the Sun) regions of reduced
star density produced by extinction in the nearest
clouds near the equator and in the Gould Belt are
seen on all plots; on plots (b), in contrast to (a) and
(c), the radial (relative to the Sun) concentrations of
stars produced by the systematic errors in $r_{rpm}$ are
clearly seen. Small regions of high star density are
clearly seen here and there -- OB associations and
their groups, for example, the associations in Orion
($Y\approx-0.2$ kpc, $Z\approx-0.15$ kpc) are seen on the $YZ$
plot.

The accuracy of the Tycho-2 photometry gives
hope that the theoretical correlation between the age
and $(B_{T}-V_{T})_{0}$ for OB stars near the MS seen in
Fig. 1 will manifest itself if we divide our sample
into three $(B_{T}-V_{T})_{0}$ subsamples:
(1) 5141 stars with $(B_{T}-V_{T})_{0}<-0.2^m$,
(2) 6561 stars with $-0.2^m<(B_{T}-V_{T})_{0}<-0.1^m$, and
(3) 8812 stars with $-0.1^m<(B_{T}-V_{T})_{0}<0^m$
(below, the subsamples are mentioned
under numbers $1-3$). The subsamples must
contain mostly stars with ages younger than 100,
$100-200$, and $200-400$ Myr, respectively, and, consequently,
must have different spatial distributions
and kinematics. However, the age of a specific star
can be estimated from its $(B_{T}-V_{T})_{0}$ only with a relative
error of 50--100\%. Therefore, here only the mean
age of the subsample stars and the age--$(B_{T}-V_{T})_{0}$
correlation make sense.

Figure 3 shows the expected large difference in the
distributions of the subsamples in projection onto the
$XY$, $XZ$, and $YZ$ planes (the designations are the
same as those in Fig. 2).

The stars older than 200 Myr form an almost flat
layer along the equator with a uniform distribution
of stars inside it (the small bend on the $XZ$ plot (c)
results from the observational selection due to extinction
in the Gould Belt), showing no evidence of the
Gould Belt, the Local Bubble, the Great Tunnel, and
other present-day structures. These structures are
clearly seen on plot (b) and especially on plot (a). In
the last 100 Myr, the layer of stars has been strongly
deformed, especially along the $Z$ axis. In fact, the
stars on plot (a) form a single 3D saddle-shaped
structure including the Gould Belt. This structure is
asymmetric relative to the Sun: the stars prevail in
the southern Galactic hemisphere, in the fourth and
especially the third quadrants (for example, the group
of associations in Orion).

These peculiarities of the distribution of young
stars are seen when analyzing the $Z$ distribution of
stars. For all subsamples, this distribution is well
fitted by a barometric law $D_{0}\cdot~e^{-|Z+Z_{0}|/Z_{H}}$, where $D_{0}$
is the density in the plane of the maximum, $Z_{0}$ is the
height of the Sun above this plane, $Z_{H}$ is the distance
from this plane at which the density decreases by a
factor of $e$ or the half-thickness of a homogeneous
layer of stars (Parenago 1954, p. 264). The BMG
fit is much poorer. The results found by the least squares
method using $r_{HIP}$ or $r_{ph}$ with the limitation
$(X^2+Y^2)^{1/2}<350$ pc for the sample to be complete
and without it are presented in the table. Note that
the number of stars with $r_{HIP}$ is appreciably smaller
than that of stars with $r_{ph}$.

\begin{table*}[!h]
\def\baselinestretch{1}\normalsize\footnotesize
\caption[]{Parameters $Z_{H}$ and $Z_{0}$ when the $Z$ distribution of stars is fitted by the barometric law.
}
\label{zz}
\[
\begin{tabular}{l|cc|cc}
\hline
\noalign{\smallskip}
   & \multicolumn{2}{c|}{$(X^2+Y^2)^{1/2}<350$} &  \multicolumn{2}{c|}{Without limitation} \\
\hline
 Subsample & $r_{HIP}$ & $r_{ph}$ & $r_{HIP}$ & $r_{ph}$ \\
\hline
\noalign{\smallskip}
 $Z_{H}$, pc: & & & & \\
 1 ($(B_{T}-V_{T})_{0}<-0.2$)       & 50 & 60 & 90 & 160 \\
 2 ($-0.2<(B_{T}-V_{T})_{0}<-0.1$)  & 60 & 55 & 75 & 70 \\
 3 ($-0.1<(B_{T}-V_{T})_{0}<0$)     & 60 & 55 & 75 & 65 \\
\hline
\noalign{\smallskip}
 $Z_{0}$, pc: & & & & \\
 1 ($(B_{T}-V_{T})_{0}<-0.2$)       & 30 & 30 & 45 & 75 \\
 2 ($-0.2<(B_{T}-V_{T})_{0}<-0.1$)  & 22 & 17 & 30 & 15 \\
 3 ($-0.1<(B_{T}-V_{T})_{0}<0$)     & 10 & 15 & 13 & 10 \\
\hline
\end{tabular}
\]
\end{table*}

The parameter $Z_{H}$ corresponds to the thickness
estimates for the layer of youngest Galactic stars, except
for $Z_{H}$ for subsample 1 without any distance limitation.
The large values of $Z_{0}$ for subsamples 1 and 2
suggest a large vertical displacement of the layer of
stars (and gas producing stars) in the last 200 Myr.
The Gould Belt results from this displacement. It
explains the value of $Z_{0}\approx25$ pc usually obtained
when analyzing young stars (Maiz-Apellaniz 2001).
$Z_{0}$ for subsample 3 ($10-15$ pc) corresponds to 13$\pm2$
found by Gontcharov (2008b, 2011) when analyzing
the spatial distribution of red giant clump and branch
stars, respectively. Thus, when calculating and mentioning
the height of the Sun above the Galactic
equator, it is necessary to specify relative to which
stars this height is considered. The height seems to
change smoothly with stellar age, because the height
of the Sun above the layer of the medium producing
stars changed with time.

A large number of regions of enhanced star density
(like the mentioned group of associations in Orion)
can be seen in Figs. 3a and 3b. The produced sample
allows these structures to be analyzed in future by
taking into account the stellar ages.

\section*{KINEMATICS}

Out of the 20 514 sample stars, 2293 stars have
radial velocities $V_r$ from the Pulkovo Compilation of
Radial Velocities (PCRV, Gontcharov 2006) and the
RAVE catalog (Steinmetz et al. 2006). Since only
eight stars from them do not enter into Hipparcos, to
avoid the influence of a few distant stars, we limited
the sample by the condition $r_{ph}<350$ pc (1954 stars
remained) or $r_{HIP}<350$ pc (1557 stars), depending
on whether $r_{ph}$ or $r_{HIP}$ was used. All our kinematic
results were obtained with both $r_{ph}$ and $r_{HIP}$.
Although a small fraction of the sample stars have
$V_r$, our subsequent conclusions about the kinematics
seem to be valid for the entire sample, because the
distributions of the stars with $V_r$, in space, in $(B_{T}-V_{T})_{0})$,
and in $M_{V_{T}}$ correspond to the distributions of all sample stars.

The distributions of the stars with $V_r$ over subsamples
1, 2, 3 are: 245, 565, 747 stars with $r_{HIP}<350$ pc and
266, 748, 940 stars with $r_{ph}<350$ pc,
respectively. However, we used the division into the
subsamples only when analyzing the distribution of
the stars in the space of velocities $UVW$. When the
velocity dispersions, components of the solar motion
to the apex, and Ogorodnikov--Milne model parameters
are calculated, modern computers allow one to
apply a moving calculation instead of the calculations
of kinematic parameters for a few subsamples and
to see their smooth changes with $(B_{T}-V_{T})_{0}$ and,
hence, with stellar age. This approach is analogous to
a moving averaging. The stars are lined up in $(B_{T}-V_{T})_{0}$
and the kinematic parameters are calculated for
200 stars with minimum $(B_{T}-V_{T})_{0}$. The star with
minimum $(B_{T}-V_{T})_{0}$ is then excluded from the set
of stars under consideration, a previously unused star
with minimum $(B_{T}-V_{T})_{0}$ is introduced instead of
it, and the calculations are repeated. As a result, we
obtain 1754 and 1357 solutions, the sets of kinematic
parameters (dispersions, components of the motion
to the apex, and Ogorodnikov--Milne model parameters),
using $r_{ph}$ and $r_{HIP}$, respectively.

In Fig. 4, the velocity dispersions (a) $\sigma(U)$,
(b) $\sigma(V)$, and (c) $\sigma(W)$ are plotted against $(B_{T}-V_{T})_{0}$.
The gray polygonal line with the step at $(B_{T}-V_{T})_{0}=-0.15^m$
indicates the dispersions adopted in
the BMG under the assumption that an age of
150 Myr corresponds to the step. The black dashed
line indicates the dependence for the initial set of
stars. The sharp peaks of the dotted line are due
to the presence of a few halo and thick-disk stars
(hot subdwarfs, blue horizontal-branch stars) and
extremely fast OB stars of an unclear nature. A star
was assigned to such runaways and was excluded
from further consideration if at least one of the
following conditions was met for it:
\begin{eqnarray*}
\label{bmg}
&\overline{U}-3\sigma(U_{BMG})<U<\overline{U}+3\sigma(U_{BMG}) \\
&\overline{V}-3\sigma(V_{BMG})<V<\overline{V}+3\sigma(V_{BMG}) \\
&\overline{W}-3\sigma(W_{BMG})<W<\overline{W}+3\sigma(W_{BMG}),
\end{eqnarray*}
where $\overline{U}$, $\overline{V}$, $\overline{W}$ are the mean values of $U$, $V$, $W$
for the corresponding $(B_{T}-V_{T})_{0}$;
$\sigma(U_{BMG})=67$, $\sigma(V_{BMG})=51$, $\sigma(W_{BMG})=42$
km s$^{-1}$ are the velocity
dispersions for the thick disk from the BMG.
In our analysis based on $r_{ph}$ and $r_{HIP}$, we excluded
61 (3\%) and 29 (2\%) runaways, respectively. The
dependence of the dispersions on $(B_{T}-V_{T})_{0}$ after the
elimination of runaways is indicated by the gray and
black curves for $r_{HIP}$ and $r_{ph}$, respectively. The lightgray
vertical lines indicate the formal accuracy of the
dispersion estimated from $r_{HIP}$ (the accuracy from
$r_{ph}$ is slightly higher). The black triangles with error
bars indicate the result from Torra et al. (2000). The
filled square with error bars indicates the result from
Dehnen and Binney (1998). Unfortunately, for the
remaining dispersion estimates encountered in the
literature, either the stellar ages cannot be established
or the results refer to a very wide range of ages. The
conclusions are the following.
\begin{itemize}
\item The results for the two types of distances agree
within the formal accuracy almost everywhere
and also agree with the results of other researchers.
\item Both in the results obtained here and in the
results of other authors, we see systematic
variations of the dispersions with $(B_{T}-V_{T})_{0}$
and, apparently, with stellar age.
\item The approximation of a stepwise change in
dispersions with age adopted in the BMG is
too rough and inaccurate, especially for the
bluest (youngest) stars. We see that the BMG
dispersions more likely correspond to those of
the initial set. The presence of an admixture
of runaways among the young stars seems to
have been disregarded in the BMG.
\item The elimination of a few runaways changed
radically the result. Therefore, it is necessary
to thoroughly study the nature of these
stars, to determine the characteristics of their
population, and to justify the criteria for their
elimination.
\item For the youngest stars ($(B_{T}-V_{T})_{0}<-0.23^m$),
the dispersions increase, confirming the previously
detected (Fig. 3 and the table) large
deformation of the layer of stars along the $Z$
axis in the last 100 Myr with the formation of
the Gould Belt.
\item The dispersion $\sigma(W)$ changes approximately
periodically. It should be checked whether this
is a real effect or an artifact.
\end{itemize}

After the elimination of runaways, let us consider
the distribution of the stars from subsamples 1--3 in
velocity space in projection onto the $UV$, $UW$, and
$VW$ planes (km s$^{-1}$). Figures 5a, 5b, and 5c for
subsamples 1, 2, and 3, respectively, show $UVW$
obtained with $r_{HIP}$ and $r_{ph}$, which give similar results.
Not only an increase in the velocity dispersions with
age, but also significant differences in the distributions
of the subsamples are clearly seen. They agree
with the analogous results from Antoja et al. (2008)
presented in their Fig. 13: plot (a) with their plot
for stars younger than 100 Myr, plot (c) with their
plot for $100-500$ Myr, and plot (b) is intermediate in
appearance. As expected, the change in the number
of stars belonging to the four main superclusters/
streams (marked in Fig. 5 by the corresponding
numbers) is most pronounced in the $UVW$ space
within the last 400 Myr:
\begin{enumerate}
\item The Pleiades ($U\approx-12$, $V\approx-22$ km s$^{-1}$; $U$, $V$, $W$
for the groups here and below were taken from Antoja
et al. (2008) and Bobylev et al. (2010)) is clearly seen
on all plots in accordance with the cluster age of about
100 Myr (Bovy and Hogg 2010).
\item Coma Berenices ($U\approx-10$, $V\approx-5$ km s$^{-1}$),
as expected, is appreciably outnumbered by the
Pleiades in the number of stars on plot (a) and is
clearly seen on (b) and (c).
\item UrsaMajor or the Sirius stream ($U\approx+9$, $V\approx+3$ km s$^{-1}$)
is absent on plot (a), manifests itself as isolated
stars on (b), and dominates on (c) in accordance with
the cluster age estimate, $350-413$ Myr (Bovy and
Hogg 2010).
\item The Hyades ($U\approx-40$, $V\approx-20$ km s$^{-1}$) manifests
itself only as isolated stars on plot (c) in accordance
with the cluster age, $488-679$ Myr (Bovy and
Hogg 2010).
\end{enumerate}

Naturally, the derived distribution differs noticeably
from that for older stars, giants (Famaey
et al. 2005), and main-sequence F and G stars
(Nordstr\"om et al. 2004); for example, the Hercules
stream ($U\approx-30$, $V\approx-50$ km s$^{-1}$) is absent here.

Let us determine the kinematics of OB stars within
the framework of the linear Ogorodnikov--Milne
model described by Gontcharov (2011). As has been
noted above, the model parameters (the solar motion
components relative to the stars $U_{\odot}$, $V_{\odot}$, $W_{\odot}$, the
partial derivatives of the velocity with respect to the
distance $M_{ux}-M_{wz}$, the Oort constants $A$, $B$, $C$, $K$,
the vertex deviation $l_{xy}$, and the angular velocity of
Galactic rotation $\Omega_{R0}$) were determined by moving
calculations, which allows their change with $(B_{T}-V_{T})_{0}$ to be traced.

In Fig. 6, the solar motion components (a) $U_{\odot}$, (b) $V_{\odot}$, (c) $W_{\odot}$
relative to the stars are plotted
against $(B_{T}-V_{T})_{0}$: just as in Fig. 4, the gray curve
for $r_{HIP}$ and the black curve for $r_{ph}$; the accuracy of
the result from $r_{HIP}$ is indicated by the light-gray vertical
lines. For comparison, the black triangles, gray
square, gray circle, black diamond, black circles, and
black square (not the popular (in references) mean
solar motion relative to MS stars of different ages
but the result for blue stars) indicate, respectively,
the results from Torra et al. (2000), Bobylev (2006),
Zabolotskikh et al. (2002), Glushkova et al. (1998),
Elias et al. (2006), and Dehnen and Binney (1998).
Here, as in our analysis of the dispersions in Fig. 4,
there is agreement of the results from the two types
of distances between themselves and with those of
other researchers and smooth systematic variations
with $(B_{T}-V_{T})_{0}$ are seen in all results.

While analyzing the discrepancies, we should note
$W_{\odot}=12$ km s$^{-1}$ from Glushkova et al. (1998). This
value is not shown in the figure, because it differs significantly
from the remaining values apparently due to
the use of only $\mu$ and/or due to the difference between
the spatial regions of the samples. There is also a clear
discrepancy between the results obtained and $U_{\odot}=12$
km s$^{-1}$ from Dehnen and Binney (1998) (the black
square on plot (a)). The use of only $\mu$ is apparently
also responsible for this discrepancy, which can give
a systematic error in the presence of radial streams
of young stars in the Galaxy. The results from Elias
et al. (2006) marked on each plot by four black circles,
two for $\mu$ and two for $V_r$, are an example of the difference
in the results when using only $\mu$ and only $V_r$: $U_{\odot}$
agree, while $V_{\odot}$ and $W_{\odot}$ disagree noticeably. It can
be suggested that $V_{\odot}$ and $W_{\odot}$ are closer to the true
value when using $V_r$ (larger values in this case) and
$\mu$ (smaller values), respectively, in accordance with
the dominant contribution of the Galactic rotation
to each velocity component. The results from Elias
et al. (2006) then agree well with our results.

For the stars with $(B_{T}-V_{T})_{0}<-0.22^m$, i.e., the
Gould Belt stars, the behavior of $U_{\odot}$ and $V_{\odot}$ changes
due to the overall motion of the Gould Belt in the
direction of the third Galactic quadrant.

In Fig. 7, the Ogorodnikov--Milne model parameters
derived here are plotted against $(B_{T}-V_{T})_{0}$. For
comparison, the triangles indicate the results from
Torra et al. (2000): we see their agreement with our
results. Unfortunately, in the remaining range of
$(B_{T}-V_{T})_{0}$, there are no other results for comparison.
Therefore, the large variations in parameters seen in
Fig. 7, especially in the range $-0.13^m<(B_{T}-V_{T})_{0}<0^m$, need to be confirmed by other studies.

The parameters $M_{wx}$ and $M_{wy}$ provide an estimate
of the accuracy of our results. They must be small,
because the OB stars retain a small thickness of
their layer over hundreds of Myr and, hence, have
no significant systematic vertical motions, with the
possible exception of sample compression/expansion
($M_{wz}$). We see from the figure that the $M_{wx}$ and $M_{wy}$
variations lie within the error band and do not differ
significantly from zero. The variations in $M_{vy}$, which
reflects the sample compression/expansion along the
$Y$ axis, are also insignificant.

The variations in the remaining parameters are
significant and occasionally correlated between themselves.
For example, at $(B_{T}-V_{T})_{0}\approx-0.13^m$, i.e.,
about 170 Myr ago, several parameters change
abruptly and since then they have changed monotonically:
the $M_{ux}$ variations show that the sample
expands after its maximum compression along $X$;
$M_{wz}$ -- periodic expansions and compressions along
$Z$ arise after 200 Myr of rest, $M_{vz}$ -- a period of
constant negative values comes after abrupt changes,
a result of the emergence of streams displaced along
$Z$ relative to the Sun and having systematic motions
along $Y$ (for example, the Pleiades),$M_{uy}$ ($=-\Omega_{R0}$) --
the Galactic rotation velocity increases, $C$ and $L_{xy}$ --
the rotation of the velocity ellipsoid, the redistribution
of the energy of horizontal stellar motions along $X$
into the energy of vertical motions along $Z$, $K$ -- the
long period of negative values is replaced by a long
period of approximately zero values.

We see that the well-known negative K-effect for
OB stars (Bobylev 2006) refers only to the stars in the
range $-0.13^m<(B_{T}-V_{T})_{0}<-0.05^m$, i.e., with ages
of about $170-300$ Myr.

\section*{CONCLUSIONS}

This study is the next step in using multicolor
broadband photometry following the selection and
analysis of the spatial distribution of stars of different
types (Gontcharov 2008a, 2008b, 2011) and
the construction of 3D maps of reddening, extinction
coefficient $R_V$ (Gontcharov 2012a), extinction
(Gontcharov 2012b), and a 3D extinction model
(Gontcharov 2009) -- here, we made an attempt to
reconcile the photometric distances, absolute magnitudes,
and extinctions for a large number of stars
of the same type. The evidence for a successful
solution of this problem includes: good agreement of
the trigonometric, photoastrometric, and photometric
distances within 350 pc of the Sun; the falling of the
stars under consideration into the region of the H--R
diagram predicted by the theory, between the zero-age
main sequence and the beginning of the subgiant
stage, after their dereddening; agreement with
the well-known spectral classification, according to
which O-A0 stars account for 94\% of the sample; the
division of the sample into three theoretically justified
subsamples by the dereddened color $(B_{T}-V_{T})_{0}$
correlating with the stellar age (less than 100, $100-200$,
and $200-400$ Myr) and the detection of expected
significant differences in the spatial distributions and
kinematics of the subsamples; the manifestation
of well-known Galactic structures in the spatial
distribution -- the Great Tunnel, the Gould Belt, the
group of associations in Orion, and others; the
manifestation of well-known streams (the Pleiades,
Coma Berenices, Ursa Major, the Hyades) in the
corresponding age subsamples; the manifestation
of a unified process in the spatial distribution and
kinematics -- the replacement of the compression of
the sample along the $X$ axis by its deformation along
the $Z$ axis $100-200$ Myr ago with the formation of
spatial structures, including the Gould Belt, outside
the equatorial plane of the Galaxy; the systematic
temporal variations in kinematic parameters of
stars (the velocity dispersions, the solar motion to
the apex, the Oort constants $A$, $B$, $C$, $K$, and the
Ogorodnikov--Milne model parameters) consistent
with the results of other authors. The detection of
temporal variations in kinematic parameters is the
main result of our study, showing that it is meaningless
to calculate these parameters for a sample of stars
with uncertain ages or with a wide range of ages.

\section*{ACKNOWLEDGMENTS}

In this study, we used results from the Hipparcos
and 2MASS projects and resources of the Strasbourg
Data Center (France), http://cds.u-strasbg.fr/. The
study was supported by Program P.21 of the Presidium
of the Russian Academy of Sciences.

\newpage

\begin{figure}
\includegraphics{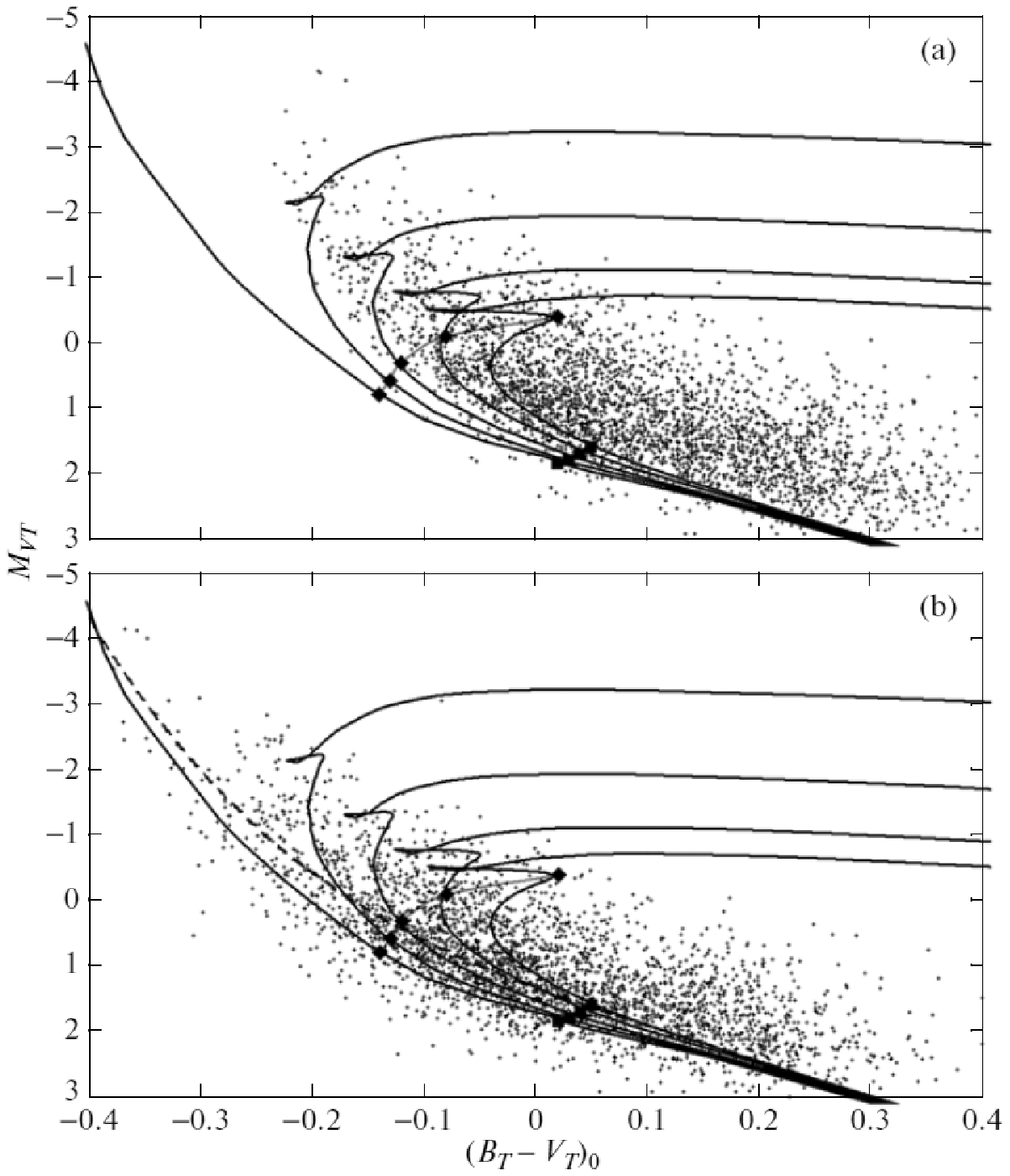}
\caption{H--R diagram. The theoretical isochrones for solar-metallicity stars with ages of 1, 100, 200, 300, and 400 Myr
are indicated by five black curves from left to right. The theoretical evolutionary tracks for solar-metallicity stars
with masses of 2.9 M$_{\odot}$ and 2 M$_{\odot}$ are indicated by the diamonds with the gray curve and the squares,
respectively. The 3237 Hipparcos stars from the preliminary sample with $\pi>5$ mas and $\sigma(\pi)/\pi<0.2$
are indicated by the crosses before (a) and after (b) their
dereddening. The adopted calibration is indicated by the black dashed line.
}
\label{oba}
\end{figure}

\begin{figure}
\includegraphics{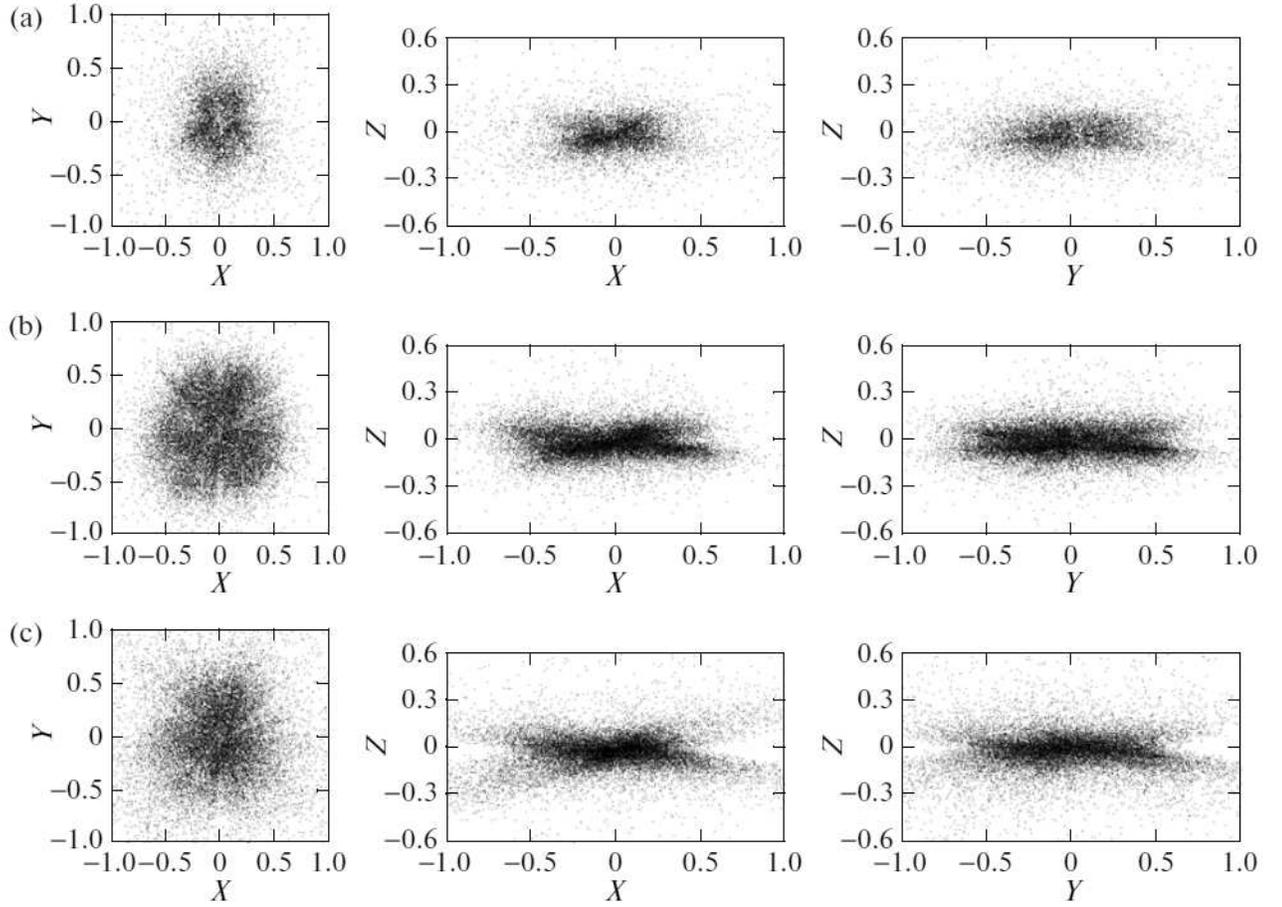}
\caption{Distribution of the sample stars in projection onto the $XY$, $XZ$, and $YZ$ planes (distances in kpc):
(a) with $r_{HIP}$ for Hipparcos stars,
(b) with $r_{rpm}$ for all 20 514 stars, and
(c) with $r_{ph}$ for them.
}
\label{xyz}
\end{figure}

\begin{figure}
\includegraphics{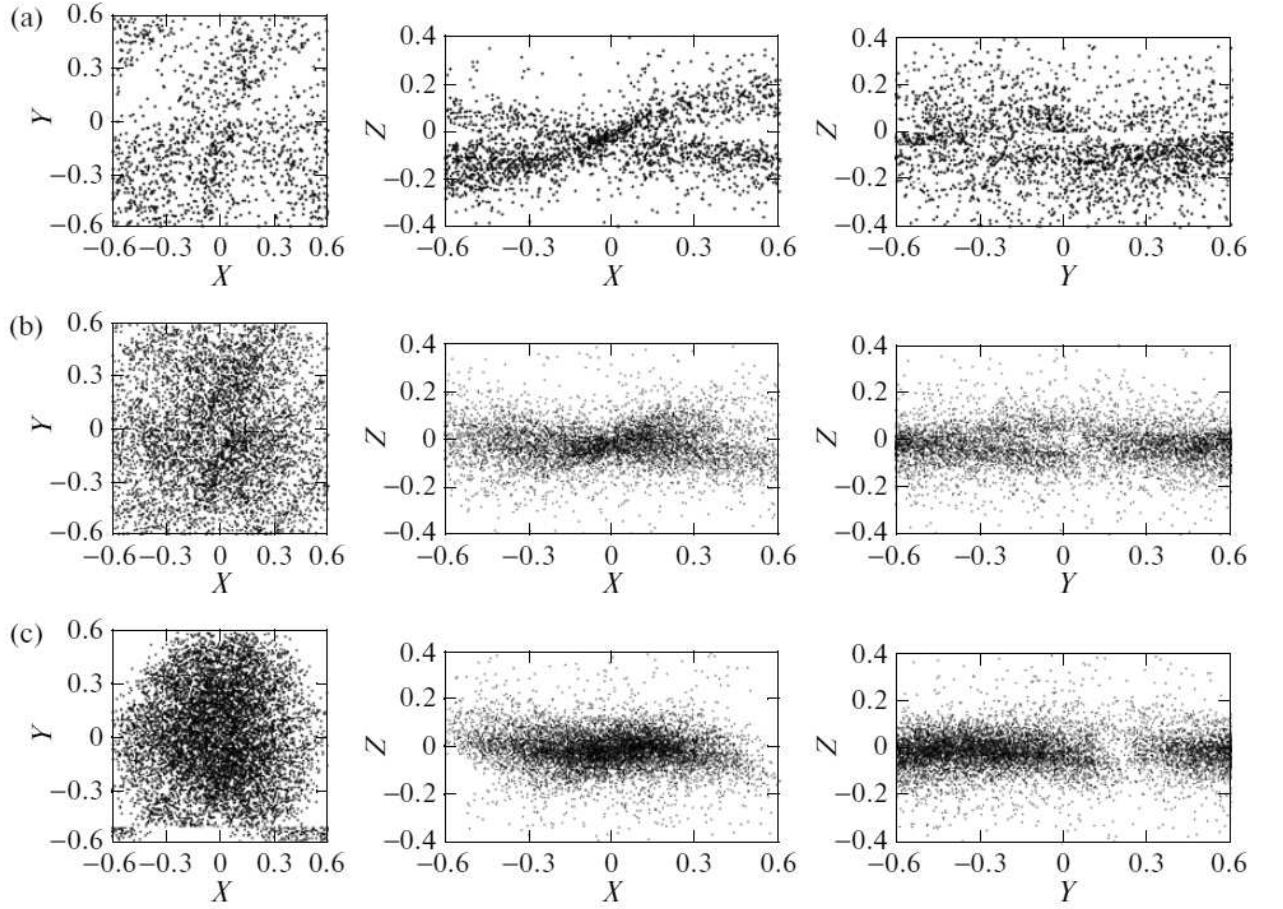}
\caption{Distribution of the stars from subsamples 1--3 in projection onto the $XY$, $XZ$, and $YZ$ planes (distances in
kpc):
(a) $(B_{T}-V_{T})_{0}<-0.2^m$ (younger than 100 Myr),
(b) $-0.2^m<(B_{T}-V_{T})_{0}<-0.1^m$ ($100-200$ Myr), and
(c) $-0.1^m<(B_{T}-V_{T})_{0}<0^m$ ($200-400$ Myr).
}
\label{xyzage}
\end{figure}

\begin{figure}
\includegraphics{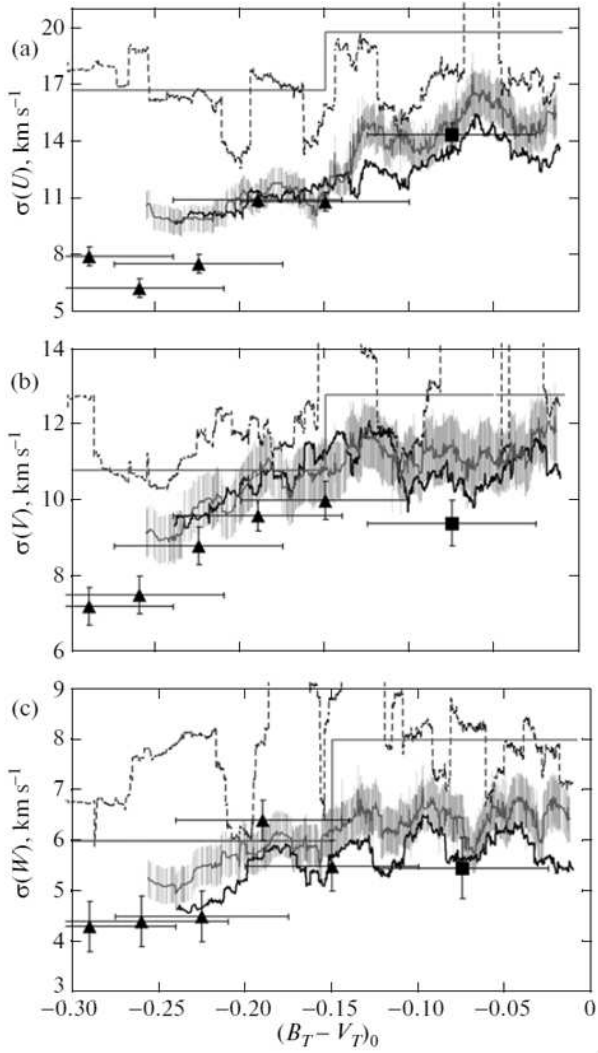}
\caption{Dependences of (a)$\sigma(U)$, (b) $\sigma(V)$, and (c) $\sigma(W)$ (km s$^{-1}$) on $(B_{T}-V_{T})_{0}$.
The dispersions from the BMG are
indicated by the gray curve with one step corresponding to an age of 150 Myr. The dependence for the initial set of stars
is indicated by the black dashed line. The dependence after the elimination of runaway stars is indicated by the gray
curve for $r_{HIP}$ and by the black curve for $r_{ph}$.
The accuracy of the result for rHIP is indicated by the light-gray vertical lines. The result from
Torra et al. (2000) is represented by the black triangles.
The result from Dehnen and Binney (1998) is represented by the black square.
}
\label{suvw}
\end{figure}

\begin{figure}
\includegraphics{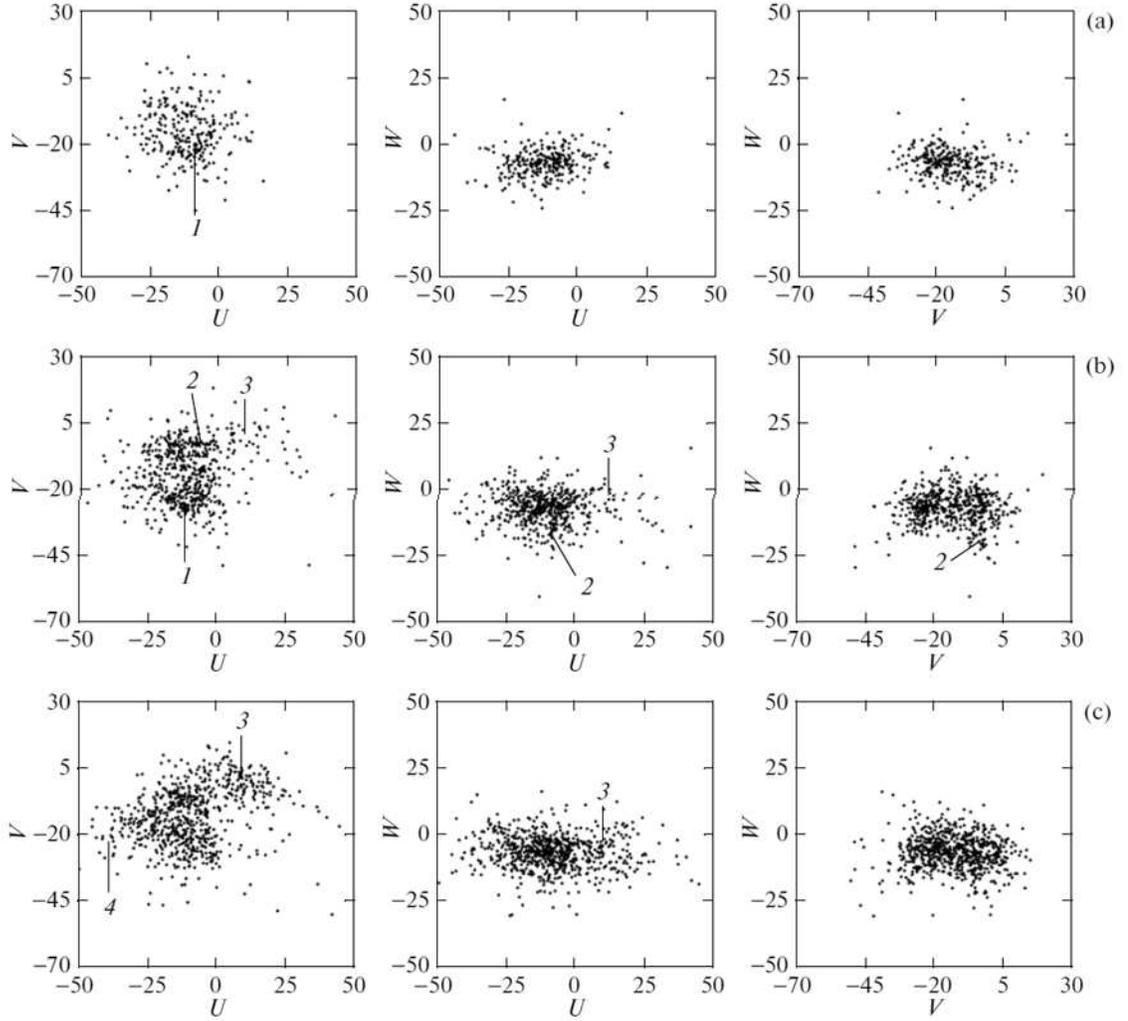}
\caption{Distribution of the subsample stars on the $UV$, $UW$, and $VW$ diagrams:
(a) 1, (b) 2, (c) 3 (velocities in km s$^{-1}$). The
numbers mark the groups discussed in the text.
}
\label{uvw}
\end{figure}

\begin{figure}
\includegraphics{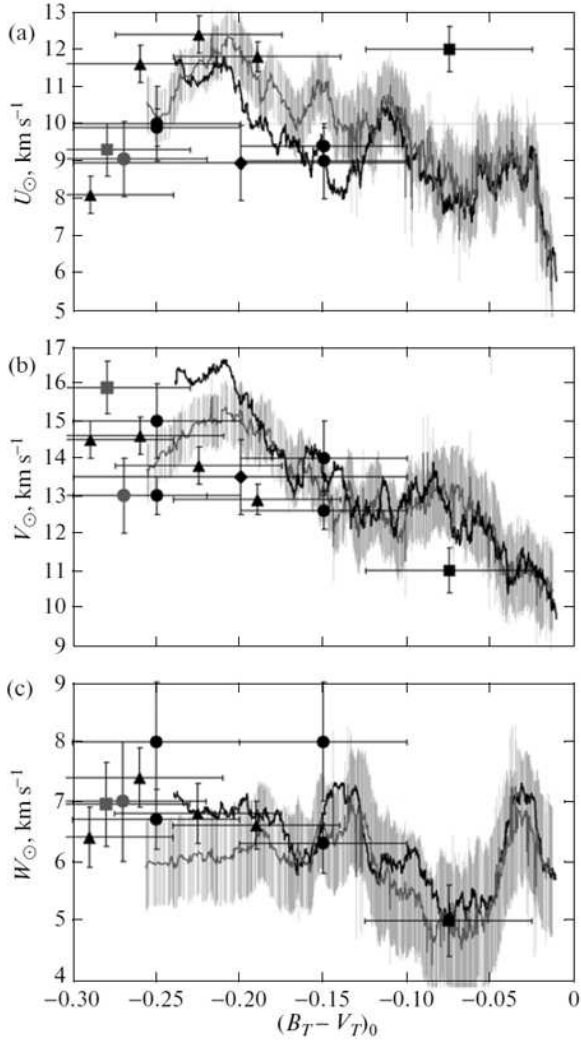}
\caption{(a) $U_{\odot}$, (b) $V_{\odot}$, and (c) $W_{\odot}$ (km s$^{-1}$) versus $(B_{T}-V_{T})_{0}$:
the gray curve for $r_{HIP}$ and the black curve for $r_{ph}$. The
accuracy of the result from $r_{HIP}$ is indicated by the light-gray vertical lines.
The results from Torra et al. (2000), Dehnen
and Binney (1998), Bobylev (2006), Zabolotskikh et al. (2002), Glushkova et al. (1998) (no $W_{\odot}$),
and Elias et al. (2006) are
indicated by the black triangles, black square, gray square, gray circle, black diamond, and black circles, respectively.
}
\label{uvwsol}
\end{figure}

\begin{figure}
\includegraphics{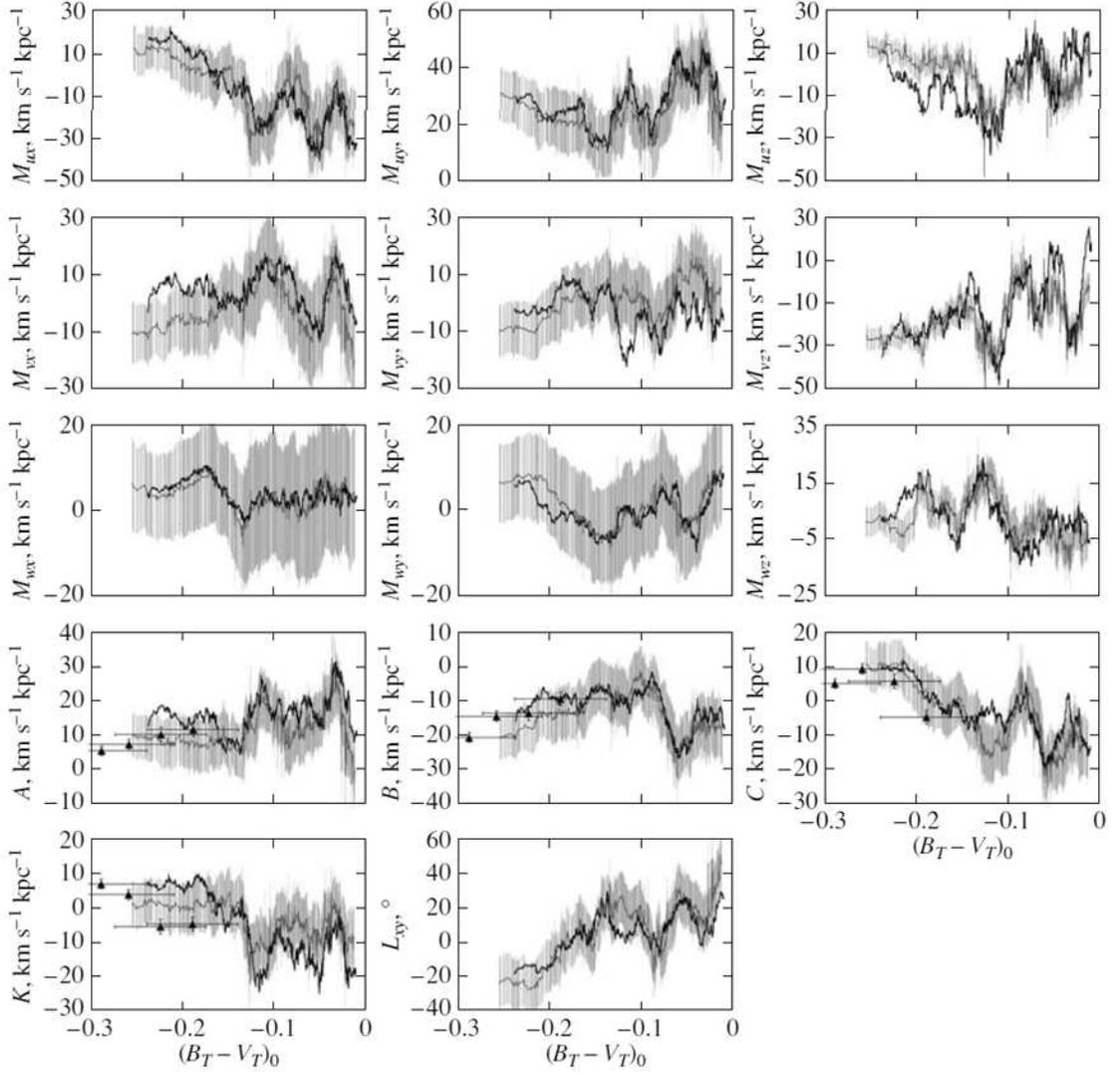}
\caption{Ogorodnikov--Milne model parameters versus $(B_{T}-V_{T})_{0}$:
the gray curve for $r_{HIP}$ and the black curve for $r_{ph}$. The
accuracy of the result from $r_{HIP}$ is indicated by the light-gray vertical lines.
The results from Torra et al. (2000) are indicated
by the black triangles.
}
\label{om}
\end{figure}

\end{document}